# Measuring Spark on AWS: A Case Study on Mining Scientific Publications with Annotation Query

Darin McBeath and Ron Daniel Jr.
Elsevier Labs
2018-01-29

## Abstract

There are many different kinds of NLP analysis that can be performed, ranging from sentence boundary detection, through parsing, and up to event extraction and beyond. In any real application, one typically needs to combine multiple types of analysis, and construct new analyses on top of the previous ones. There are several methods for doing that combination; one particularly interesting method is through a query language based on region algebras[1][2], such as GATE's Mímir[3][4]. Unfortunately, Mímir has issues with scaling to tens of millions of documents, which is the requirement for our purposes. We have been using Apache Spark[5], and particularly the Databricks implementation, as the basis for analyzing millions of scientific articles. That kind of operation raises the question of how to get the best balance of cost and performance.

To answer that question, we developed a benchmark task based on NLP and using a tool we call Annotation Query (AQ). AQ is a program that provides the ability to query many different types of NLP annotations on a text, as well as the original content and structure of the text. The query results may provide new annotations, or they may select subsets of the content and annotations for deeper processing. Like Mímir, AQ is based on region algebras. Our AQ is implemented to run on a Spark cluster. In this paper we look at how AQ's runtimes are affected by the size of the collection, the number of nodes in the cluster, the type of node, and the characteristics of the queries. Cluster size, of course, makes a large difference on performance so long as skew can be avoided. We find that there is minimal difference in performance when persisting annotations serialized to local SSD drives as opposed to deserialized into local memory. We also find that if the number of nodes is kept constant, then a storage optimized configuration perform the best. But if we factor in total cost, the compute optimized configuration provides the best performance relative to cost.

## 1 Introduction

Many groups are working to analyze the STM (Scientific, Technical, and Medical) literature, as evidenced by the series of Workshops on Mining Scientific Publications. They may wish to extract important facts, to build knowledge bases that can be used to provide better medical care, to provide faster development of new drugs or other research outcomes, or to help practitioners make better use of the overabundance of published research that exceeds human capacities to monitor. Several years ago, Elsevier Labs began creation of a Content Analytics Toolbench (CAT) to make analysis of our content faster, easier, and more productive[6]. CAT is in its third major iteration and is now based on the implementation of Apache Spark. CAT3 has preserved the main architectural point from the original CAT – to process the content and store the stand-off annotations as structured text files on S3[7]. Others can use those files without repeating expensive processing. Instead they can concentrate on using those stored annotations to help in making new analyses. Layers of annotation build up over time so that sophisticated operations will rarely require large amounts of new code, but instead are built on the work of others.

A key tool that simplifies this vision is Annotation Query (AQ)[1]. It allows existing text and annotations to be read, and easily queried. The results may be used to directly create new annotations, or at least to filter out a subset of the original annotations as input to some more sophisticated function that may provide new annotations. In one example, a member of Elsevier Labs generated an early layer of annotations that provided sentence breaks, POS (Part Of Speech) tags, and noun phrase chunks. A second person in the group used that information to create Open Information Extraction triples using a basic implementation of the ReVerb method[8]. A third member took the noun phrases from the ends of the extractions and matched them to concept names in a medical ontology, providing a smaller but higher-quality set of open relations. A fourth member combined those open relations with the medical ontology using a method known as Universal Schemas[9]. This produced a medical knowledge base which could be used to find new extensions to the ontology based on items found in the literature[10].

The contributions of this paper are:
1) Introduction of AQ, an open source software package for querying stand-off annotations.
2) Analysis of AQ performance with different queries , collections of annotations, and machine configurations.
3) Key observations and conclusions from the analysis of AQ performance.

This paper is organized as follows: It begins by describing the Annotation Query tool, which can be run on a single machine, or at scale on a Spark cluster to process large amounts of content (such as the millions of articles and chapters in Elsevier's archive). The paper then goes on to look at how its execution time is affected by the number of documents and annotations, the complexity of the query, the number of nodes in the cluster, and the kinds of machine used in the cluster. Section 2 provides background information on previous search tools that inspired Annotation Query. Section 3 describes the Annotations used, the various operators available in the query language, and the architecture of the system. Section 4 describes the procedure followed to characterize AQ's capabilities and performance. Section 5 provides the results, and Section 6 the discussion.  Lastly Section 7 offers our conclusions and Section 8 some notes on future work.

## 2 Related Background
AQ was inspired in large part by the work on region algebras[1][2] and the query language in the Mimir[4] tool from GATE[3]. These allow one to form regions over the document's string of characters by specifying the start and end point – using the typical convention of indexing the gap between characters so that end-start gives the length of the region. Zero length regions are allowed but negative lengths are not. The algebras then specify a number of useful operators, such as CONTAINS, CONTAINED_IN, BEFORE, AFTER, etc. These operators take lists of regions as inputs and return a filtered list of regions as outputs. When augmented with text or regex search to create regions from text, and import tools to create regions corresponding to document structure or natural language syntax, a useful query language is created.

---
[1] AQ is released as open source as is available at http://github.com/elsevierlabs-os/AnnotationQuery

As an example, assume we want to investigate whether citations that appear in the Methods section of papers have different characteristics than those in the Introduction section. We could use queries[2] like:

```
M = (<sentence> CONTAINING <citation>)
   CONTAINED_IN
    (<section> CONTAINING
     (<section_name> CONTAINING "Method"))

I = (<sentence> CONTAINING <citation>)
   CONTAINED_IN
    (<section> CONTAINING
     (<section_name> CONTAINING "Introduction"))
```

We have now isolated those two populations of sentences and can proceed to examine them for distinguishing characteristics.

A few other annotation search tools have been created, of which Mimir[4] from the GATE[3] project may be the best known. Mimir and GATE's annotation format were strong influences on AQ and the CAT annotation format. Unlike Mimir, AQ does not build an index of all the text. This harms its search speed for strings, but it makes quick iteration and stacking of annotations much easier.

## 3 System Description

### 3.1 System Architecture

Annotation Query has a simple architecture. It runs on a Spark cluster and uses whatever degree of parallelization is available in the cluster. AQ is implemented in Scala. All the annotations are held in Datasets of AQAnnotation (a class that is described next). Query operations are Scala functions that typically take two `Dataset[AQAnnotation]` as input arguments and return one as the result. This makes it easy to compose the functions to obtain more elaborate queries. The results, being Datasets, can be saved to disk at any time, typically as Parquet files[11]. As Datasets, they fit in cleanly with the rest of the Spark infrastructure. The outputs of AQ can be used by feature extraction and other functions in MLlib and other Spark libraries.

---

[2] These queries are expressed in a pseudo query language for simplicity. The syntax for actual queries used in the study is provided later in Table 3.

## 3.2 Annotation Format

As described in the section on background research, our annotations identify regions in the original text according to the document ID, and the character offsets of the start and end of the region[3]. Zero-length regions are permitted. In addition to the three elements for the region, each annotation also includes fields for a unique identifier; an annotation type name (such as sentence, token, ce:para) that indicates the main meaning of the annotation; an annotation set name which functions like an XML namespace and allows some annotation type names to be reused (e.g. genia^sentence and scnlp^sentence as examples for the Genia[12] and Stanford Core NLP[13] libraries we use. Finally, a variable length Map of name-value pairs is also provided to convey additional information. For example, the genia^word annotation needs properties for the original text of the token, for the lemma of the token, for its part of speech tag, etc. The scnlp^dep (Stanford CoreNLP dependency parse) annotation has different properties. AQAnnotations are defined as a Scala class. That definition, and an example instance, are shown below in Listing 1.

**Listing 1: AQAnnotation class definition and example instance**

```
AQAnnotation(docId: String,        // Document Id
        annotSet: String,          // Annotation set (eg. scnlp, ge)
       annotType: String,          // Annotation type (e.g  token, sentence)
     startOffset: Long,            // Starting offset into source file
       endOffset: Long,            // Ending offset into source file
         annotId: Long,            // Annotation Id unique within file/Dataset
      properties: Option[scala.collection.Map[String,String]] = None) // Properties

AQAnnotation(123456789, // DocID
         ge,         // GENIA annotation set – basic NLP for science documents
         NP,         // Noun Phrase annotation type
         1439,       // starting offset in document
         1450,       // ending offset in document
         376,        // unique ID of annot in Dataset or Parquet file
         Some(Map("orig" -> "other trees", // Original text
                  "pos" -> "jj nns")       // Parts of speech (adj, plural noun)
```

## Annotation Query Functions

After the Parquet files of AQAnnotation are loaded by Spark into one or more Spark Dataset[AQAnnotation], they can be provided as arguments to the functions listed in Table 1. Spark SQL is used to implement the underlying logic for the functions. Since the functions return a Dataset[AQAnnotation], complex queries can be constructed through the composition of these atomic building block functions. In the descriptions below, A ,B, and C represent a Spark Dataset[AQAnnotation].

---

[3] To be precise, when the original document is XML, the annotation position is specified by offsets into the XML document's string value, rather than the XML file. This avoids a host of difficulties, the explanation of which is outside the scope of this paper.

*Table 1: Query Functions*

| | |
|---|---|
| **FilterSet(**A,*setName*) | Filter annotations in A to only those of the named annotation set. |
| **FilterType(**A,*typeName*) | Filter annotations in A to only those of the named annotation type. |
| **FilterProperty(**A, *name, value*) | Filter annotations in A to only those that contain a property with the specified name and value. |
| **RegexProperty(**A, *name, regex*) | Filter annotations in A to only those that contain a property with the specified name and a value matching the specified regex. |
| **Contains(**A,B) | Return annotations from A whose range covers any annotation in B. |
| **ContainedIn(**A,B) | Return annotations from A whose range is covered by any annotation in B. |
| **Before(**A,B) | Return annotations from A whose range is before any annotations in B. |
| **After(**A,B) | Return annotations from A whose range is after any annotations in B. |
| **Between(**C,A,B) | Return annotations from C whose range is after any annotations in A and before any annotations in B. |
| **Sequence(**A,B) | Return new annotation (encompassing range of A and B) where A range is before any annotations in B. |
| **MatchProperty(**A,B,*name*) | Filter annotations in A whose value for a property with the specified name matches the same property name with the same value in B. |
| **Preceding(**A,B**)** | For each annotation in B, return the annotation and an Array[AQAnnotation] from A that immediately precede the annotation. This differs from the Before function in that the relationships for each annotation in B and the annotations from A that occur before each annotation in B are maintained. |
| **Following(**A,B**)** | For each annotation in B, return the annotation and an Array[AQAnnotation] from A that immediately follow the annotation. This differs from the After function in that the relationships for each annotation in B and the annotations from A that occur after each annotation in B are maintained. |

# 4 Method

To analyze AQ's performance on different dataset sizes, three collections (small, medium, and large) of journal article and book chapter XML files were created. Each collection also contained AQAnnotations for two annotation sets. The first set, Original Markup, provides the elements from the original document's XML markup as regions. For simplicity in benchmarking, the various attributes of the XML elements were not copied into the Map field in each AQAnnotation, although that is normally done. The second annotation set, Genia[12], provides annotations for sentences, noun phrases, verb phrases, and individual tokens. All of those are regions as well. Here again, most name-value pairs were not loaded into the Map, only the single 'orig' attribute that holds the text of the original token.

Table 2 reports the sizes of the three collections in terms of the number of documents, the total number of annotations, the size of the annotations (stored as snappy compressed Parquet files), and the size of the Dataset[AQAnnotation] when serialized to disk. Note that the number of documents differ between om and genia. This is because we only apply genia annotations to documents with a publication date > 2000. The parquet file is larger than the serialized AQAnnotations since we are only populating a couple of values in the property map and we are leveraging the Spark built-in encoders for the serialized AQAnnotations.

*Table 2: Document Collection Characteristics*

| Collection | # Documents | # Annotations | Parquet File Size | AQAnnotation Size |
|---|---:|---:|---:|---:|
| Small (om) | 43,354 | 169,480,491 | 2.9 GB | 1.1 GB |
| Small (genia) | 40,928 | 289,850,996 | 8.5 GB | 3.5 GB |
| Medium (om) | 1,100,790 | 1,734,613,441 | 28.4 GB | 10.7 GB |
| Medium (genia) | 815,027 | 2,797,395,400 | 79.4 GB | 33.8 GB |
| Large (om) | 15,302,153 | 21,796,090,428 | 361.1 GB | 136.5 GB |
| Large (genia) | 7,947,730 | 38,205,933,831 | 1098.8 GB | 461.5 GB |

Table 3 provides the queries that were used to evaluate performance. Queries test retrieval of both rare and very common annotation types, rare and common property values, rare and common regular expressions applied to property values, and several combinations of queries that use the Contains/ContainedIn operations. The number of annotations returned for each query in the small, medium and large collection is defined in the annotation collection count column. The queries were executed through notebooks on a Databricks[4] runtime 3.3 AWS cluster running Spark 2.2 with default settings.

---

[4] Databricks is a company founded by the creators of Apache Spark, that aims to help clients with cloud-based big data processing using Spark. Databricks grew out of the AMPLab project at University of California, Berkeley that was involved in making Apache Spark, a distributed computing framework built atop Scala. Databricks is the main committer to the open source Spark, but they also have proprietary implementations of some operations for higher performance in their for-fee product.

*Table 3: Tested Queries*

| Query Name | Annotation Collection Count | Query String |
|---|---|---|
| OM Type Common | Small: 63982<br>Medium: 888419<br>Large: 11450929 | FilterType(om, "ce:abstract") |
| Genia Type Common | Small: 7289076<br>Medium: 74769571<br>Large: 1023026736 | FilterType(genia, "sentence") |
| Genia Type Very Common | Small: 210931654<br>Medium: 2029330974<br>Large: 27765916569 | FilterType(genia, "word") |
| Genia Attribute Very Common | Small: 13054093<br>Medium: 109747913<br>Large: 1665482184 | FilterProperty(genia, "orig","the") |
| Genia Attribute Typical | Small: 15048<br>Medium: 421566<br>Large: 3568493 | FilterProperty(genia, "orig","heart") |
| Genia Attribute Rare | Small: 60<br>Medium: 1047<br>Large: 35001 | FilterProperty(genia, "orig","adrenocortical") |
| Genia Attribute Regex Common | Small: 5867694<br>Medium: 59325352<br>Large: 807984311 | RegexProperty(genia, "orig","^he*") |
| Genia Attribute Regex Rare | Small: 415<br>Medium: 7361<br>Large: 174468 | RegexProperty(genia, "orig","^adrenoc*") |
| Sentence over Heart | Small: 12775<br>Medium: 362790<br>Large: 3074530 | Contains(FilterType(om, "ce:sentence"),<br>    FilterProperty(genia,"orig","heart")) |
| Heart in Sentence | Small: 15048<br>Medium: 421566<br>Large: 3568493 | ContainedIn(FilterProperty(genia,"orig","heart"),<br>    FilterType(genia,"sentence")) |
| Sentence in Abstract | Small: 345569<br>Medium: 4615525<br>Large: 48406955 | ContainedIn(FilterType(genia,"sentence"),<br>    FilterType(om,"ce:abstract")) |
| Heart in Sentence in Abstract | Small: 932<br>Medium: 28917<br>Large: 186305 | ContainedIn(FilterProperty(genia,"orig","heart"),<br>    ContainedIn(FilterType(om, "ce:sentence"),<br>        FilterType(om, "ce:abstract"))) |

We focused our testing on memory optimized (r3 class machines), compute optimized (c3 class machines) and storage optimized (i3 class machines) AWS EC2[14] configurations. To initialize the system, the Dataset[AQAnnotation] for both the OM (Original Markup) and the Genia annotations were loaded from Parquet files on S3, hash partitioned on docId and stored in memory and/or serialized to local drives[5].  We also disabled the DBIO caching layer of Databricks by specifying spark.conf.set("spark.databricks.io.cache.enabled", "false").  For memory optimized configurations, Dataset[AQAnnotation] was persisted to a combination of memory and disk.  For the compute and storage optimized configurations, Dataset[AQAnnotation] was persisted to disk.  Before running any of the queries, we verified that the Datasets were 100% persisted. For all of the storage and computed optimized configurations, the Datasets were persisted 100% to the local SSD drives. For the memory optimized configuration, the Datasets were roughly persisted 90% to memory and 10% to disk. Each query was run 3 times on each collection using the memory, compute, and storage cluster configurations. Between each run the cluster was restarted.

---

[5] A key observation is that AQAnnotations do not cross document boundaries. By partitioning on the document ID, we guarantee that all the annotations for a document will be placed in the same partition. This greatly reduces the need for shuffles and is key to getting acceptable performance.

The compute, storage, and memory optimized configuration details for the small, medium, and large collections are recorded below in Tables 4a-c. The value for cores, RAM, disk, and cost are per node.

*Table 4a: Small Collection Cluster Configurations*

| Configuration | Instance Type | Nodes | Cores | RAM | Disk | Partitions | Cost/hour |
|---|---|---|---|---|---|---|---|
| Memory Optimized | r3.2xlarge | 2 | 8 | 61 GB | 160 GB SSD | 32 | $0.665 |
| Compute Optimized | c3.2xlarge | 2 | 8 | 15 GB | 2 x 80 GB SSD | 32 | $0.42 |
| Storage Optimized | i3.2xlarge | 2 | 8 | 61 GB | 1.9 TB SSD | 32 | $0.624 |

*Table 4b: Medium Collection Cluster Configurations*

| Configuration | Instance Type | Nodes | Cores | RAM | Disk | Partitions | Cost/hour |
|---|---|---|---|---|---|---|---|
| Memory Optimized | r3.4xlarge | 4 | 16 | 122 GB | 320 GB SSD | 256 | $1.33 |
| Compute Optimized | c3.4xlarge | 4 | 16 | 30 GB | 2 x 160 GB SSD | 256 | $0.84 |
| Storage Optimized | i3.4xlarge | 4 | 16 | 122 GB | 2 x 1.9 TB SSD | 256 | $1.248 |

*Table 4c: Large Collection Cluster Configurations*

| Configuration | Instance Type | Nodes | Cores | RAM | Disk | Partitions | Cost/hour |
|---|---|---|---|---|---|---|---|
| Memory Optimized | r3.8xlarge | 8 | 32 | 244 GB | 2 x 320 GB SSD | 1024 | $2.66 |
| Compute Optimized | c3.8xlarge | 8 | 32 | 60 GB | 2 x 320 GB SSD | 1024 | $1.68 |
| Storage Optimized | i3.8xlarge | 8 | 32 | 244 GB | 4 x 1.9 TB SSD | 1024 | $2.496 |

We did additional testing with the medium collection to evaluate the impact of different number of nodes while keeping all other variables constant and the impact of different number of partitions while keeping all other variables constant. Those configuration details are recorded below in Table 4d-e. Once again, the value for cores, RAM, disk and cost are per node.

*Table 4d: Medium Collection Cluster Configurations (Vary Number Nodes)*

| Configuration | Instance Type | Nodes | Cores | RAM | Disk | Partitions | Cost/hour |
|---|---|---|---|---|---|---|---|
| Memory Optimized | r3.4xlarge | 4/8/16 | 16 | 122 GB | 320 GB SSD | 256 | $1.33 |
| Compute Optimized | c3.4xlarge | 4/8/16 | 16 | 30 GB | 2 x 160 GB SSD | 256 | $0.84 |
| Storage Optimized | i3.4xlarge | 4/8/16 | 16 | 122 GB | 2 x 1.9 TB SSD | 256 | $1.248 |

*Table 4e: Medium Collection Cluster Configurations (Vary Number Partitions)*

| Configuration | Instance Type | Nodes | Cores | RAM | Disk | Partitions | Cost/hour |
|---|---|---|---|---|---|---|---|
| Memory Optimized | r3.8xlarge | 8 | 32 | 244 GB | 2 x 320 GB SSD | 256/512/1024 | $2.66 |
| Compute Optimized | c3.8xlarge | 8 | 32 | 60 GB | 2 x 320 GB SSD | 256/512/1024 | $1.68 |
| Storage Optimized | i3.8xlarge | 8 | 32 | 244 GB | 4 x 1.9 TB SSD | 256/512/1024 | $2.496 |

## 5 Results

The following section documents the results from our testing. Unless otherwise specified, when looking at the graphs, the y-axis represents the median elapsed time (in seconds) for each query name on the x-axis.

Readers will note that the query times are typically in the range of a few seconds to a few minutes, instead of the sub-second range. The reasons for this are provided in the Discussion section below.

Elapsed time per run

We first evaluated the elapsed time for each run of the same query on each configuration. The times between runs were very similar. The small collection set and compute optimized configuration results provided in Figure 1 are representative of what we witnessed (other than the elapsed times getting bigger as the collection size increased, of course).

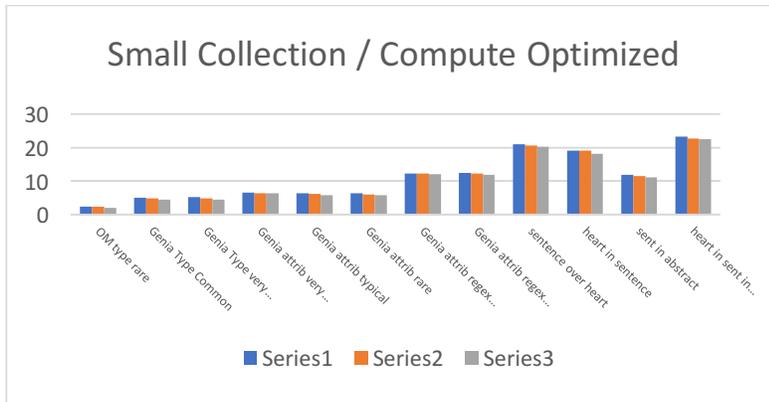

*Figure 1 – Median elapsed time for queries on Small Collection and Compute Optimized configuration.*

Comparing narrow operations

Looking at the small collection as a representative example, Table 5a provides the median elapsed times for the queries across the compute, storage, and memory optimized configurations. Other collections and configurations exhibited similar behavior. We specifically wanted to test different query functions such as type, attribute, and attribute regex and how these are impacted by the number of annotations returned for the query. These types of query operations are commonly referred to as narrow operations.

*Table 5: Small Collection Narrow Operation Performance*

| Query Name | Compute Optimized | Storage Optimized | Memory Optimized |
|---|---|---|---|
| Genia Type Common | 4.94 | 4.17 | 2.15 |
| Genia Type Very Common | 4.94 | 4.32 | 2.29 |
| Genia Attribute Very Common | 6.4 | 5.75 | 3.83 |
| Genia Attribute Typical | 6.21 | 5.43 | 3.43 |
| Genia Attribute Rare | 6.04 | 5.5 | 3.27 |
| Genia Attribute Regex Common | 12.32 | 11.2 | 9.58 |
| Genia Attribute Regex Rare | 12.17 | 11.17 | 9.63 |

Comparing compute, storage, and optimized configurations

Using the configurations outlined in Table 4a-c, we compared the performance of different instance types using the small, medium, and large collections. Figures 5-7 provide the median elapsed times when running the queries against the 3 different collections for the 3 different configurations.

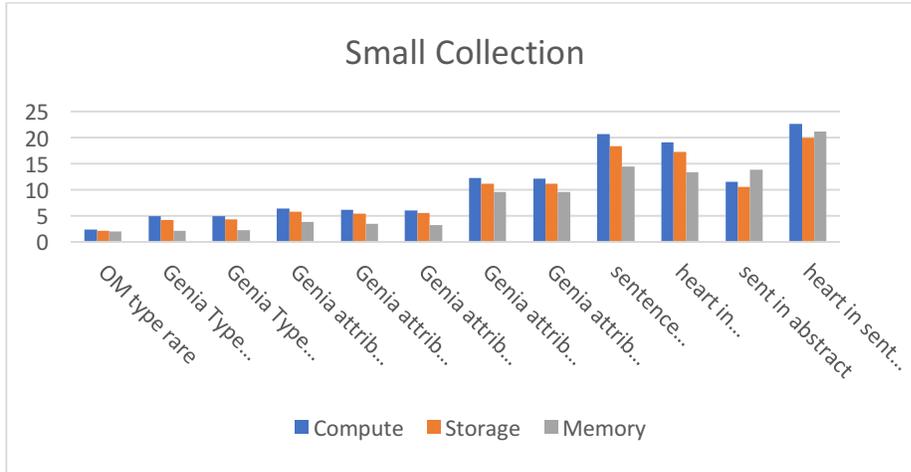

*Figure 2 - Median elapsed time on Small Collection for Compute, Storage, and Memory Optimized configurations.*

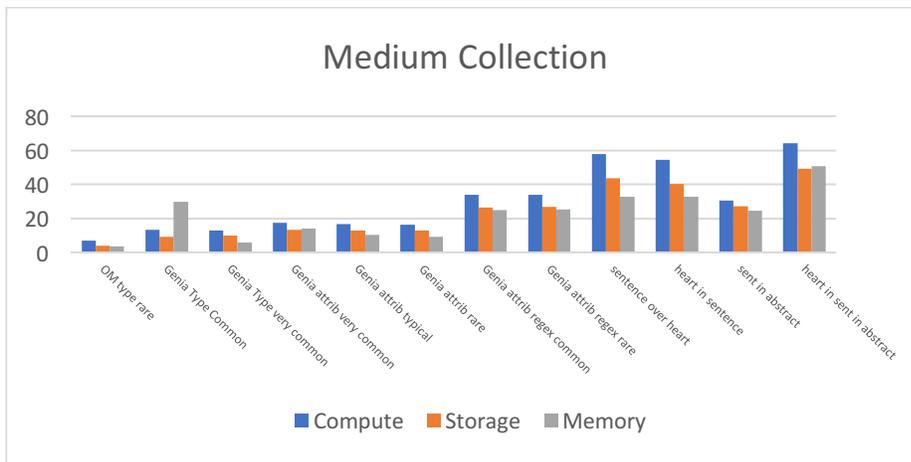

*Figure 3 - Median elapsed time on Medium Collection for Compute, Storage, and Memory Optimized configurations.*

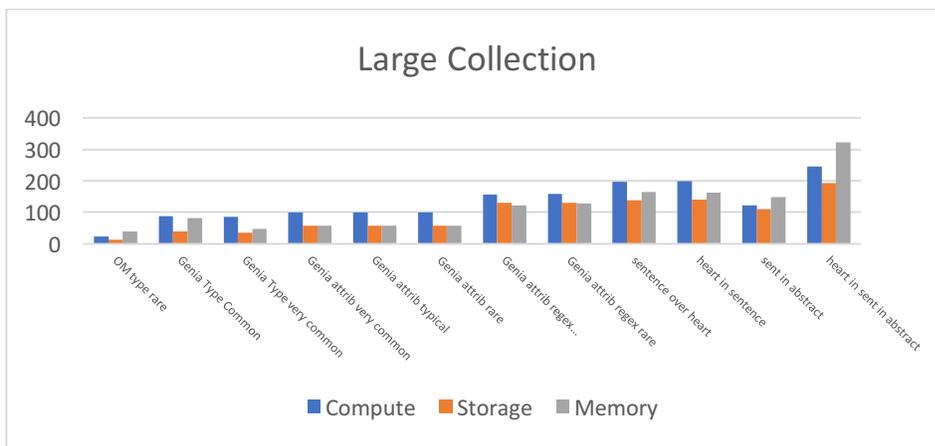

*Figure 4 - Median elapsed time on Large Collection for Compute, Storage, and Memory Optimized configurations.*

Varying the number of nodes

Using the configuration outlined in Table 4d, we looked at changing the number of nodes while keeping all other variables constant. Figures 5-7 provide the median elapsed times when running the queries against the medium document set for the compute, storage, and memory optimized configurations when using 4, 8, and 16 node clusters.

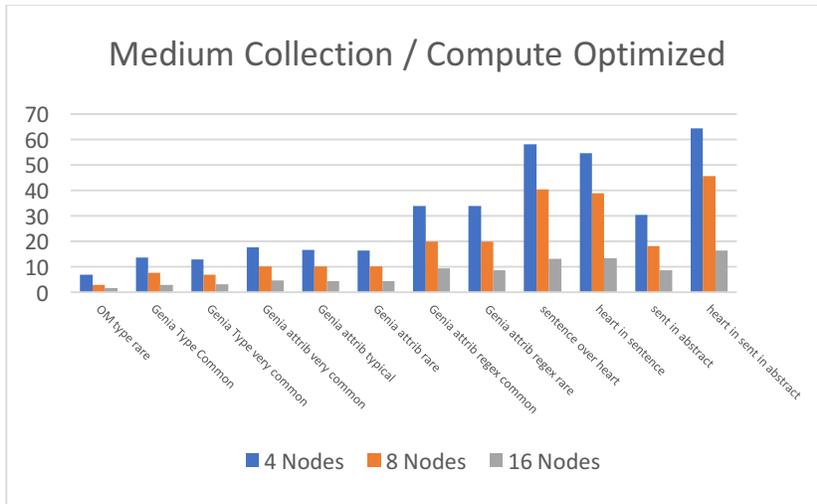

*Figure 5 – Median response on Medium Collection and Compute Optimized configuration (4, 8, and 16 nodes).*

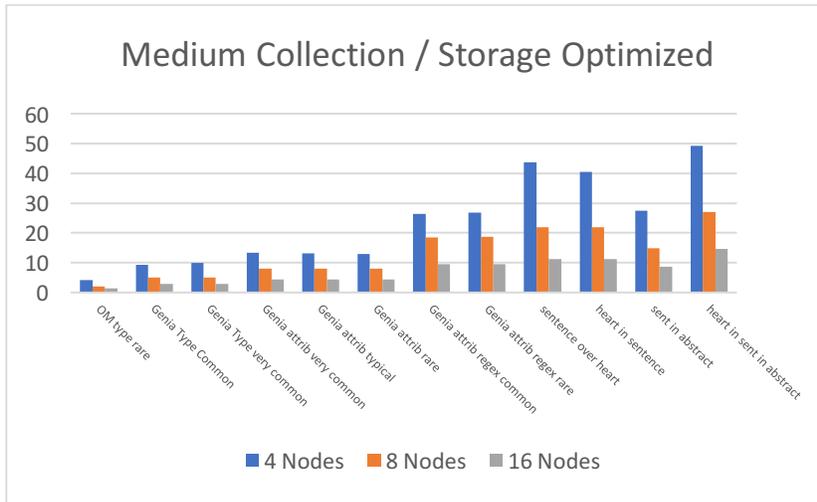

*Figure 6 - Median elapsed time on Medium Collection and Storage Optimized configuration (4, 8, and 16 nodes).*

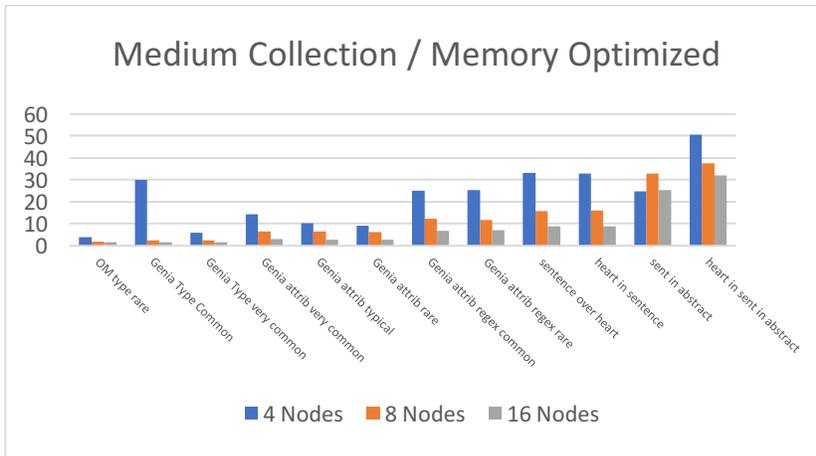

*Figure 7 - Median elapsed time on Medium Collection and Storage Optimized configuration (4, 8, and 16 nodes).*

Varying the number of partitions

Using the configuration outlined in Table 4e, we looked at changing the number of partitions while keeping all other variables constant. Figures 8-10 describe the median elapsed times when running the queries against the medium document set for the compute, storage, and memory optimized configurations when using 256, 512, and 1024 partitions.

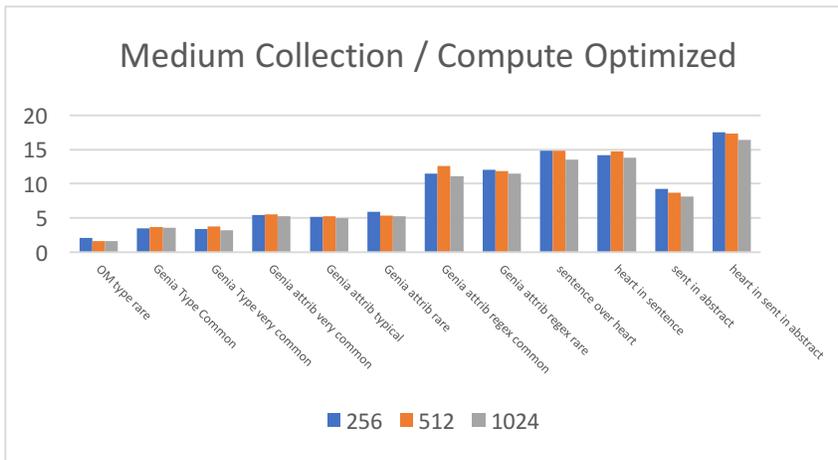

*Figure 8 - Median elapsed time on Medium Collection and Compute Optimized configuration (256, 512, and 1024 partitions).*

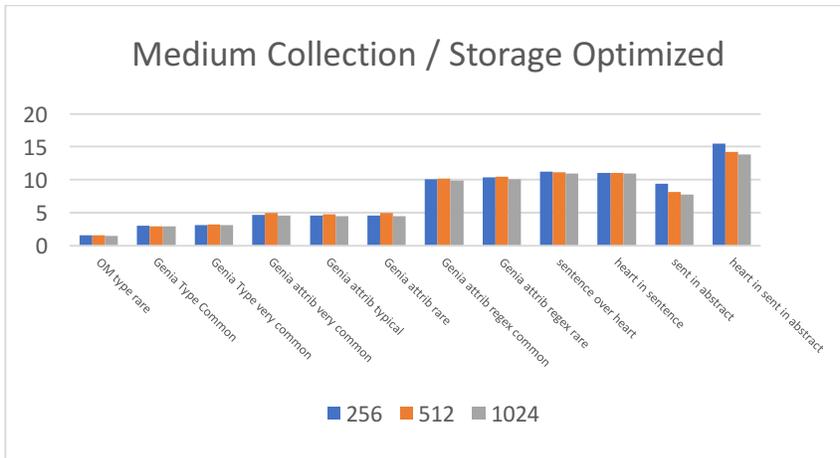

*Figure 9 - Median elapsed time on Medium Collection and Storage Optimized configuration (256, 512, and 1024 partitions).*

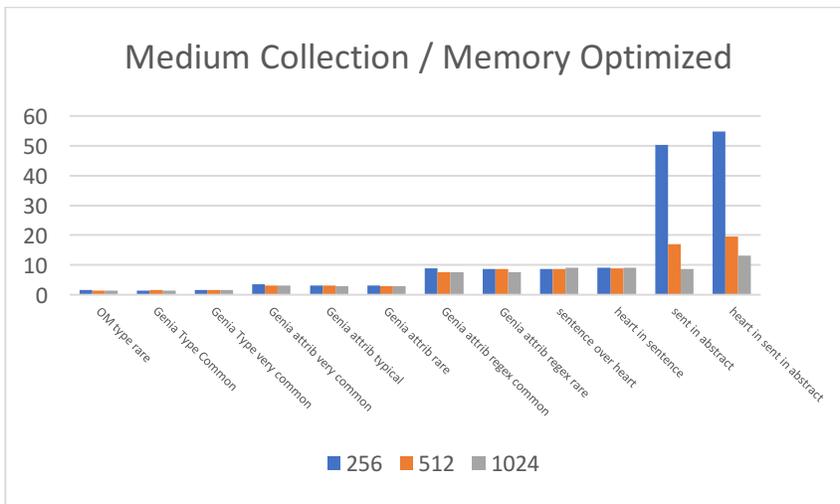

*Figure 10 - Median elapsed time on Medium Collection and Memory Optimized configuration (256, 512, and 1024 partitions).*

Equivalent Total Cost

Using the configuration outlined in Table 4d, we looked at doubling the number of nodes for the compute optimized configuration. Since the cost for a compute optimized node is roughly 35% the cost of the storage and memory optimized configurations, doubling the number of nodes for the compute optimized configuration (while keeping the number of nodes unchanged in the storage and memory optimized configurations) provides a comparison based on similar total cost across the configurations. Figures 11-12 describe the median elapsed times when running the queries against the medium document set for the compute, storage, and memory optimized configurations when doubling the number of nodes in the compute optimized configuration.

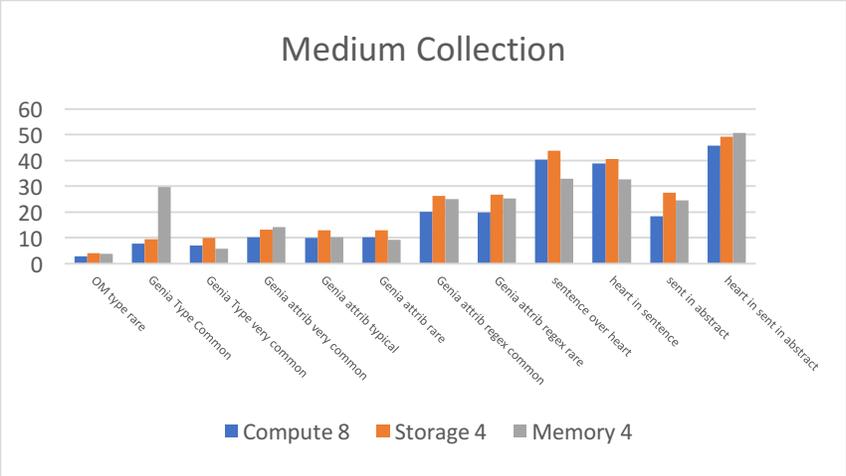

*Figure 11 - Median elapsed time on Medium Collection with 2x nodes for Compute Optimized configuration.*

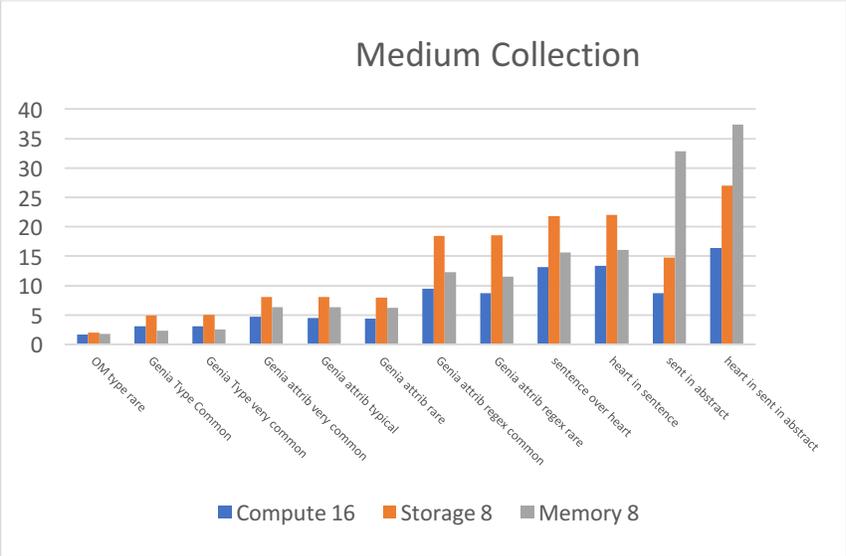

*Figure 12 - Median elapsed time on Medium Collection with 2x nodes for Compute Optimized configuration.*

## 6 Discussion

We used a variety of queries, collection sizes, and cluster configurations to analyze AQ performance.

When looking at AQ narrow operations, one thing we noticed was the elapsed time for common and very common types were nearly identical within a given configuration and collection of data. This also held true for queries across very common, typical and rare attributes. We did see increases when going from the Type column to an attribute field in the Map, and going from constant string queries to regex queries. But the frequency of the value had little impact. This makes sense because all of these queries are essentially full-table scans and independent of the number of results returned.

When comparing the compute, storage, and memory optimized configurations, we noticed minimal difference in performance between the storage and memory optimized configurations. This also indicates there is not a significant difference in performance between storing the annotations serialized to SSD as opposed to deserialized into memory.  Similar observations were reported in a separate study comparing RDD and DataFrame performance[15].  When viewing the Spark UI, we did notice that garbage collection could contribute up to 20% of the overall elapsed time for the compute and storage optimized configurations while there was minimal garbage collection overhead for the memory optimized configuration. As we increase the size of the cluster by adding additional nodes and increasing the number of partitions, we will want to closely monitor garbage collection to make sure it doesn't become a larger portion of the elapsed time. While the compute optimized configuration typically has the worst performance, we must balance this with the fact that it costs 35% less to run this configuration.  When looking at the results, our clusters were undersized and would have benefited from additional nodes and partitions.  This is specifically pronounced in the testing of the medium and large collections.  Our additional testing on the medium collection for increasing the number of nodes and partitions confirms this assumption.

When increasing the number of nodes, the compute and storage optimized configuration behave as expected.  The elapsed time decreased by roughly 50% each time the number of nodes was doubled.  Of course, the number of partitions will need to be at least (and preferably 2x to 4x) the number of cores available on the cluster to achieve this reduction. The memory optimized configuration exhibited quite a bit of variability in elapsed times and the elapsed time decrease did not follow the same trend.  In particular, for the queries requiring nested joins (containedIn), some of the partitions were heavily skewed which made performance for the memory optimized configuration slower than both the compute and storage optimized configurations.  Since the partitioning was the same for all configurations, it's unclear why only the memory optimized configuration experienced skewing. We plan to further investigate this anomaly.

When increasing the number of partitions, the compute and storage optimized configuration behave as expected.  The elapsed time stayed roughly the same as the number of partitions increased.  This was expected since we did not increase the size of the cluster (number of nodes) but only changed the number of partitions.  The memory optimized configuration behaved similarly with the exception of the nested joins.  Because of the skewing noticed previously, as we increased the number of partitions the elapsed time also decreased.

In all of the cases, our query times were in the range of a few seconds to a few minutes, rather than the typical goal of sub-second response. There are several reasons for this. First, there is no requirement that the query be run interactively. Once an algorithm to create new annotations from existing ones has been developed, it can be run as a batch job on whatever cluster size is most cost-effective. Second, the long times are associated with the largest collection size. During algorithm development, a small collection is typically adequate and gives faster turnaround. The medium and large collections would be used for testing and production but not development.  Third, we are using very small clusters for the amount of data. The small cluster, for example, is only two nodes. Doubling the number of nodes would cut the query time nearly in half if a user is looking for interactive performance. But for long-running and non-interactive use, we have to balance the cost of machine time on a long-running, lightly loaded application vs. the startup overhead of reading in large numbers of annotations into on-demand clusters.

# 7 Conclusion

The following are our conclusions from the evaluation of AQ performance.

1) AQ elapsed response time is independent of the number of annotations returned for a given narrow operation.
2) AQ can achieve equivalent elapsed response times when serializing Datasets to AWS Storage Optimized configuration (i3) and deserializing Datasets to Memory Optimized configuration (r3).
3) Increasing the number of nodes in the cluster can decrease elapsed AQ response time if there are enough partitions to take advantage of all the executors and there is no skewing.
4) Increasing the number of partitions can decrease elapsed AQ response time if there is skewing.
5) With all other variables the same, a given AQ nested join query can behave differently between AWS Storage Optimized configuration (i3) and Memory Optimized configuration (r3).
6) AQ performance was the most consistent and predictable for the Compute Optimized and Storage Optimized configurations. AQ performance was more variable and unpredictable with the Memory Optimized configuration.

With similar performance and price points to the Memory Optimized configuration, it is advantageous to use the Storage Optimized configuration as more data can be stored on the locally attached SSD drives. The Compute Optimized configuration is a cost-effective alternative as these machines are typically 35% the cost of the equivalent storage or memory optimized configurations. While performance is typically slower for compute optimized configurations when using the same number of nodes as memory and storage optimized configurations, the performance can be improved and even superior to storage or memory optimized configurations by doubling the number of compute optimized nodes. Although the number of nodes is doubled, the total cost between this configuration will be roughly equivalent to a storage or memory optimized configuration with half the number of nodes. The downside of the Compute Optimized configuration is the limited amount of disk space.

# 8 Future Work

In addition to the version of AQ that runs on Spark, and operates on `Dataset[AQAnnotation]`, a standalone version of AQ has been developed that can run on single machines without Spark SQL. It operates on `Array[AQAnnotation]`. Query operations are unaffected and code written for one version ports almost transparently to the other. Using this approach, very small machines can be leveraged with the annotations for each document processed independently. For certain use cases, this approach will likely be a more cost effective and performant option.

We are considering alternatives to the current Map collection for the properties in the AQAnnotation.  While the current approach provides flexibility, the implementation currently pays a price in both storage and performance.  One option currently being investigated are the fastutil extensions for the Java Collections framework.   By decreasing the memory footprint for `Dataset[AQAnnotation]`, we could consider persisting these datasets as deserialized objects in memory and further boost performance.  This also has the potential to reduce our garbage collection times when  restoring serialized AQAnnotations.

We are currently leveraging the default parameter settings for Spark.  While these settings are likely fine for many scenarios, we could potentially further boost performance by adjusting some of these parameters.  For example, we could explore the impacts for adjusting the spark.memory.fraction and spark.memory.storage.fraction settings.

We are currently leveraging the default Spark SQL Catalyst optimizer (which is primarily rule based optimizations).  With Spark 2.2, an option for cost based optimization was introduced. We plan on exploring cost based optimization for AQ but will need to think through the impacts of our Map collection for the properties.

While we currently hash partition on docId, skewing still negatively impacts some queries.   We hope to explore other partitioning strategies and perhaps develop a custom partitioner to minimize skewing and improve overall performance.

The current implementation for text query is a linear scan of the orig property in the annotations. We plan to investigate using  Succinct on Apache Spark to determine whether a suffix array index will significantly improve performance for text queries.

At this time the query functions are only provided in Scala. Porting them to Python waits on demand and should be straightforward.

[8] A. Fader, "Identifying Relations for Open Information Extraction," in *Conference on Empirical Methods in Natural Language Processing, Proceedings of the Conference*, 2011.

[9] S. Reidel, "Relation extraction with matrix factorization and universal schemas," in *2013 Conference of the North American Chapter of the Association for Computational Linguistics: Human Language Technologies, Proceedings of the Main Conference*, Atlanta, 2013.

[10] P. Groth, "Applying Universal Schemas for Domain Specific Ontology Expansion," in *5th Workshop on Automated Knowledge Base Construction (AKBC) 2016*, San Diego, 2016.

[11] Apache Software Foundation, "Parquet," [Online]. Available: https://parquet.apache.org/.

[12] Genia Project, "Genia," [Online]. Available: http://www.geniaproject.org/.

[13] Stanford CoreNLP, "Stanford CoreNLP – Natural Language Software," [Online]. Available: https://stanfordnlp.github.io/CoreNLP/.

[14] Amazon Web Services, "Amazon EC2," [Online]. Available: https://aws.amazon.com/ec2.

[15] K. Zhang, " Understanding and Improving Disk-based Intermediate Data Caching in Spark," in IEEE International Conference on Big Data , Proceedings of the Conference, 201719